\documentclass[preprint,showpacs,preprintnumbers,amsmath,aps,prd]{revtex4-1} 
\usepackage{graphicx}

\begin{document}
\title{Holographic Entanglement Entropy and hidden Fermi Surfaces}
\author{ZhongYing Fan}\email{zhyingfan@gmail.com}
\affiliation{Department of Physics, Beijing Normal University, Beijing 100875, China}
\date{\today}

\begin{abstract}
We prove that the purely classical gravity dual of Fermi and non-Fermi liquids exist by employing the logarithmic behavior of entanglement entropy to probe Fermi surfaces. For isotropic systems, the logarithmic behavior originates only from the deep UV region of the minimal area surface. For anisotropic systems, the surfaces' configuration becomes largely modified by spatial anisotropy and series of solutions exist. By imposing the null energy condition we show that the specific heat behaves as $C\propto T^\alpha$ where $\alpha\leq1$, in both systems. In the end, we also
 present an effective gravity model for anisotropic background. However, the anisotropic scaling solutions with logarithm violation haven't been covered in this model.
\end{abstract}
\maketitle

\section{Introduction}
Recent years, AdS/CFT correspondence has been widely used to study strongly coupled fermionic systems\cite{2,3,4,5,6,7}. It was shown that holography can describe quantum liquids, especially non-Fermi liquids which still lacks of a proper description in condensed-matter theory (CMT). By adding probe fermions in the bulk, people can explicitly calculate the boundary retarded Green function $G_R(\omega,k)$ and analytically show the existence of Fermi surfaces \cite{4,5,6,7}. However, the number of Fermi surfaces are only of order $O(1)$ in this procedure and usually the Luttinger relation which equates the total charge density to the volumes enclosed by the Fermi surfaces was badly violated \cite{4,5,8}, except for the electron star background \cite{6,7}. What's more, the bulk fermions only correspond to the gauge-invariant operators in the boundary and thus the dual Fermi surfaces are only of these composite operators. Up to now, One can't directly manipulate the Green function and detect the Fermi surfaces of boundary elementary fermionic operators in holography \cite{9,10}. Thus we call these Fermi surfaces hidden.
Interestingly, in a recent paper, N.Ogawa et al have proposed an elegant approach to this conundrum \cite{1}. They define systems with Fermi surfaces by requiring that their entanglement entropy has the logarithmic violation of the area law. The number of Fermi surfaces is expected of order $O(N^2)$  and the Luttinger relation can be partly rescued in some way \cite{10}.\\
For $d$ spatial dimensional systems, a strip shape subsystem $A$ is defined by:
 \begin{equation} A=\{(x_1,x_2,...,x_d)|-\frac{\ell}{2}\leq x_1 \leq \frac{\ell}{2},0\leq x_2,x_3,...,x_d \leq L\} \end{equation}
 When the size $\ell$ of the subsystem $A$ is large enough, the entanglement entropy $S_A$ will be substantially modified and behaves like \cite{1}:
\begin{equation} S_A=\gamma \frac{L^{d-1}}{\epsilon^{d-1}}+\eta L^{d-1} k_F^{d-1} log(\ell k_F)+O(\ell^0) \label{2}\end{equation}
where $\gamma$ and $\eta$ are numerical constants, and $\eta$ is positive, $\epsilon$ is the UV cut off, $k_F$ is the Fermi momentum or the average of Fermi momentums when many Fermi surfaces exist. On the other hand, in holography, the entanglement entropy is given by \cite{11,12,13}
\begin{equation} S_A=\frac{Area(\gamma_A)}{4 G_N} \label{fsa}\end{equation}
where $G_N$ is the bulk Newton's constant and $\gamma_A$ is the minimal area surface which coincides with $\partial A $ at the boundary. With the assumption that the logarithmic behavior of the entanglement entropy $S_A$ originates from the IR contribution of the minimal area surface, Ogawa et al show that only non-Fermi liquids have classical gravity duals. By imposing null energy conditions which leads to additional constraints on the metric, they show that the specific heat eventually behaves like $C\propto T^\alpha, \mbox{with}\ \alpha\leq 2/3\ (d=2)\ \mbox{and}\ \alpha\leq 3/5\ (d=3)$. Thus, only part of non-Fermi liquids can be constructed in this way. The standard Fermi liquids ($\alpha=1$) and some other non-Fermi liquids are not allowed, possibly having no purely classical gravity duals. Is this true? \\
Since working in the IR geometry means it is simply to extract the IR pieces of the full minimal area surface. It is possible that the other omitted regions is the dominant part of the surface area and leads to the logarithmic behavior of the entanglement entropy\cite{16}. Motivated by this idea, we will explicitly show that it indeed works sufficiently for general isotropic systems with $d$ spatial dimension when the deep UV part of the surface is dominant ($d\geq 2$) in this paper. The index of specific heat will be extended to $\alpha\leq 1$ and thus include Fermi and all of non-Fermi liquids.

On the other hand, since the realistic systems in the boundary are generally anisotropic, people have recently constructed anisotropic black branes \cite{15} in holography to further study the properties of the boundary systems. From this perspective, it is also important to investigate the hidden Fermi surface information and confirm the existence of the purely classical gravity dual of Fermi and non-Fermi liquids in the anisotropic gravity background. We find that the configuration of the minimal surface is controlled by two functions product so that many series of solutions which corresponds to either IR or UV piece of the minimal area surface dominant case are allowed. The Fermi and non-Fermi liquids are also naturally allowed in this case. Finally, we also present a classical gravity dual for anisotropic systems.
\section{Isotropic systems}
For self consistency, we first briefly demonstrate the calculation process given in Ref\cite{1}. To be general, we consider $d+1$ dimensional boundary systems with Fermi surfaces which is dual to $d+2$ dimensional gravity backgrounds. The metric is taken to be
\begin{equation} ds^2=\frac{R^2}{r^2} (-f(r)dt^2+g(r)dr^2+dx_i^2)\label{metric} \end{equation}
where $R$ is AdS radius; $i=1,2,...,d$. Note that this metric preserves translational and rotational symmetry and thus is dual to isotropic systems in the boundary. We also require this metric is asymptotically $AdS_{d+2}$, so we have
\begin{equation} f(0)=g(0)=1 \end{equation}
We are only interested in the strip subsystem $A$ (1). For other shape subsystems, one can refer to Ref\cite{1,9}. The minimal area surface $\gamma_A$ can be specified by the surface $x_1=x_1(r)$. Without loss of generality, we take the Newton's constant $G_N=1/4$ in the following. According to eq.(\ref{fsa}), the holographic entanglement entropy is found to be
\begin{equation} S_A=Area(\gamma_A)=2^{d-1} R^d L^{d-1} \int_\epsilon^{r_*} \frac{\mathrm{d} r}{r^d} \sqrt{g(r)+x_1'(r)^2} \end{equation}
where $r_*$ is the turning point which leads to $x_1'$ divergent. $\epsilon$ is the UV cut off. Since $\gamma_A$ is the minimal area surface, the variational principle for $x_1(r)$ gives rise to
\begin{equation} x_1'(r)=\frac{r^d}{r_*^d} \sqrt{\frac{g(r)}{1-\frac{r^{2d}}{r_*^{2d}}}} \end{equation}
Thus, we obtain the width $\ell$ and the entanglement entropy $S_A$ as follows
 \begin{equation} \ell=2 \int_0^{r_*} \mathrm{d} r \frac{r^d}{r_*^d} \sqrt{\frac{g(r)}{1-\frac{r^{2d}}{r_*^{2d}}}} \label{ell} \end{equation}
 \begin{equation} S_A=2^{d-1} R^d L^{d-1} \int_\epsilon^{r_*} \frac{\mathrm{d} r}{r^d} \sqrt{\frac{g(r)}{1-\frac{r^{2d}}{r_*^2d}}}\label{sa} \end{equation}
 Obviously, the relation between the entanglement entropy $S_A$ and the width $\ell$ is controlled only by the function g(r). Assume that the IR piece of the minimal surface $\gamma_A$ is dominant, we can set $g(r)$ as
  \begin{eqnarray}
  g(r)& \simeq & (\frac{r}{r_F})^{2n} \qquad (r\gg r_F) \nonumber \\
      & \simeq & 1 \qquad \qquad (r\ll r_F) \label{gr}
   \end{eqnarray}
 where $r_F$ is a certain scale which is undetermined. Substitute eq.(\ref{gr}) into eq.(\ref{ell}) and (\ref{sa}), we find
 \begin{equation} \ell\sim c_n \frac{r_*^{n+1}}{r_F^n}; \label{iell}\end{equation}
 \begin{equation}
 S_A  \sim \gamma_d \frac{L^{d-1}}{\epsilon^{d-1}}+d_n R^d L^{d-1} \frac{r_*^{n-(d-1)}}{r_F^n} \label{isasa}
 \end{equation}
where $c_n\ \mbox{and}\ d_n$ are positive constants and $\gamma_d=2^{d-1} R^d/(d-1)$. Note that in the very IR limit, we have $r_*\gg r_F$, or equally $\ell \gg r_F$. The eq. (\ref{iell}) and (\ref{isasa}) lead to
 \begin{equation} S_A=\gamma_d \frac{L^{d-1}}{\epsilon^{d-1}}+k_n \frac{R^d L^{d-1}}{r_F^{d-1}} (\frac{\ell}{r_F})^{\frac{n-(d-1)}{n+1}}+O(\ell^0) \end{equation}
 The first term is divergent and agrees with the area law which is expected from asymptotically $AdS$ background. The second term is the leading finite part of $S_A$ which depends on $\ell$ as a pow law. An exception occurs when $n=d-1$. In this case, the behavior of $S_A$ is indeed obtained as
  \begin{equation} S_A=\gamma_d \frac{L^{d-1}}{\epsilon^{d-1}}+k_{d-1}\frac{R^d L^{d-1}}{r_F^{d-1}} log(\frac{\ell}{r_F})+O(\ell^0) \label{log} \end{equation}
 This is the logarithmic behavior we expected for the existence of Fermi surfaces. Compare eq.(\ref{log}) with eq.(2), the scale parameter $r_F$ can now be interpreted as the Fermi level $\sim k_F^{-1}$ or the average of Fermi levels when many Fermi surfaces exist. \\
 Up to now, we have only fixed the function $g(r)$ by requiring the logarithmic behavior of the finite part of the entanglement entropy while the tt-component of the metric (\ref{metric}) i.e. the function $f(r)$ isn't involved in. To make sure what quantum liquids we have in this background, an additional and physically sensible condition is needed.
 For this purpose, let's impose the null energy conditions
 \begin{equation} T_{\mu\nu}N^\mu N^\nu\geq 0 \label{null} \end{equation}
 where $T_{\mu\nu}$ denotes the energy stress tensor of matter fields; $N^\mu$ is any null vector. In the absence of a specific matter field, $T_{\mu\nu}$ can be calculated from Einstein's equations
 \begin{equation} R_{\mu\nu}-\frac 12 g_{\mu\nu} R =2\pi T_{\mu\nu} \label{tuv} \end{equation}
 For simplicity, the null vector $N^\mu$ can be chosen as
 \begin{equation} N^t=\frac{1}{\sqrt{-g_{tt}}}, \quad N^r=\frac{\cos\theta}{\sqrt{g_{rr}}}, \quad N^{x_1}=\frac{\sin\theta}{\sqrt{g_{x_1x_1}}} \label{nu}\end{equation}
 where $\theta$ is an arbitrary constant. Obviously, equations (\ref{null})-(\ref{nu}) apply to both isotropic and anisotropic systems. Focus on the IR geometry and assume
 \begin{equation} f(r)\sim r^{-2m},\quad g(r)\sim r^{2n} \label{fg}\end{equation}
 For isotropic systems (\ref{metric}), the above conditions lead to
 \begin{equation} m\geq n\  \mathrm{and}\ m\geq 0\label{mn}\end{equation}
 In the end, we found the behavior of the specific heat as\cite{1}
 \begin{equation} C \propto T^\alpha \label{c}\end{equation}
 \begin{equation} \alpha  = \frac{d}{m+n+1} \label{alpha}\end{equation}
 To show the logarithmic behavior of $S_A$, we need $n=d-1$. Therefore
 \begin{equation} \alpha \leq \frac{d}{2d-1} \label{ialpha}\end{equation}
 Note $d\geq2$. When $d=2, \alpha\leq 2/3$ and $d=3, \alpha \leq 3/5$ which is precisely consistent with \cite{1}. Clearly, eq.(\ref{ialpha}) contains only a portion of non-Fermi liquids. The standard Landau-Fermi liquids and other non-Fermi liquids are not included in this procedure. To generalize above results, one needs to consider the contribution of the deep UV region, instead of the deep IR piece, of the minimal area surfaces. We start from a new assumption
\begin{equation} g(r)= 1+\frac{r^{2n}}{r_F^{2n}} \label{grgr}\end{equation}
Note that in the deep interior $r\gg r_F$, eq.(\ref{grgr}) reduces to eq.(\ref{gr}). Set $\Lambda =r_*/r_F \gg 1$ and $u=r/r_*$. Substitute eq.(\ref{grgr}) into eq.(\ref{ell}) and (\ref{sa}), we get
\begin{eqnarray}
\ell/r_F &=& 2 \Lambda \int_0^1 \mathrm{d}u u^d \sqrt{\frac{1+(\Lambda u)^{2n}}{1-u^{2d}}} \nonumber \\
                 & \sim & c_n \Lambda^{\beta}+...  \label{newell}
\end{eqnarray}
where the strip length is assumed to behave as $\ell/r_F \sim c_n \Lambda^{\beta}$. $c_n$ is a numerical constant, and $\beta$ is a positive constant related to $n$ in some way. The omitted terms is of order $O(\Lambda^{\beta-1})$ which seems to be rather bad. Fortunately, $\log{(\ell/r_F)}\sim {c_n} \log{\Lambda}+O(\Lambda^0)$ which is what we truly need to estimate the finite part of the entanglement entropy. In the same way, $S_A$ is obtained as follows
\begin{eqnarray}
S_A &=& \frac{2^{d-1} R^d L^{d-1}}{r_*^{d-1}} \int_{\epsilon/r_*}^1 \frac{\mathrm{d}u}{u^d} \sqrt{\frac{1+(\Lambda u)^{2n}}{1-u^{2d}}} \label{bsasa} \\
    & \approx & \frac{2^{d-1} R^d L^{d-1}}{r_*^{d-1}} \int_{\epsilon/r_*}^{r_F/r_*} \frac{\mathrm{d}u}{u^d} \sqrt{1+(\Lambda u)^{2n}} \label{sasa}
\end{eqnarray}
where $u^{2d}$ term in the integral has been omitted in the second line with a appropriate cut off at $r=r_F$ for the entropy integral. Clearly, the $u^{2d}$ term is dominant when $r \rightarrow r_*$ which locates in the deep IR region. Since we have assumed that the deep UV region contributes to the leading finite part of the entanglement entropy $S_A$, the dominant scale is actually far away from $r_*$. So we expect (\ref{sasa}) is sensible. We will show this point explicitly in the following. \\
From eq.(\ref{sasa}) and by simple calculation (see Appendix), one find if $n=(d-1)/2$, we will have
\begin{equation}
S_A= \gamma_d \frac{R^d L^{d-1}}{\epsilon^{d-1}}+d_n \frac{R^d L^{d-1}}{r_F^{d-1}} \log{\Lambda}+... \label{ssa}
\end{equation}
which agrees with eq.(\ref{log}) perfectly. We emphasize that the dominant scale for this behavior to emerge is given by $r_{UV}=r_F/\Lambda^{\delta}\ll r_F$, thus locating in the deep UV region with the UV cut-off scale $\epsilon$ fixed, where $\delta$ is some positive constant to ensure this relation. Now let's estimate the leading finite corrections $\delta (S_A)$ given by $u^{2d}$ term. When $r \rightarrow r_*$, eq.(\ref{grgr}) reduces to eq.(\ref{gr}), eq.(\ref{bsasa}) reduces to eq.(\ref{isasa}). Therefore
\begin{equation} \delta (S_A) \sim  \frac{R^d L^{d-1}}{r_F^{d-1}} \Lambda^{-( \frac{d-1}{2})}\end{equation}
Thus, for any $d$ ($d\geq2$), $\delta{S_A}/(S_A)_{fin} \ll 1$, where $(S_A)_{fin}$ denotes the finite part of $S_A$, the $u^{2d}$ term's effect is indeed negligible. \\
Combining these results, we finally obtain
\begin{equation} S_A= \gamma_d \frac{R^d L^{d-1}}{\epsilon^{d-1}}+e_n \frac{R^d L^{d-1}}{r_F^{d-1}} \log{\frac{\ell}{r_F}}+O(\ell^0) \label{endsa}  \end{equation}
From conditions (\ref{null})-(\ref{alpha}) and now $n=(d-1)/2$, we find the specific heat behaves like
\begin{equation} C\propto T^\alpha,\quad \alpha \leq 1 \end{equation}
Thus in this case both the Fermi liquids ($\alpha=1$) and all the non-Fermi liquids ($\alpha < 1$) can be constructed.
\begin{figure}[tbp]{
\includegraphics[width=7.5cm]{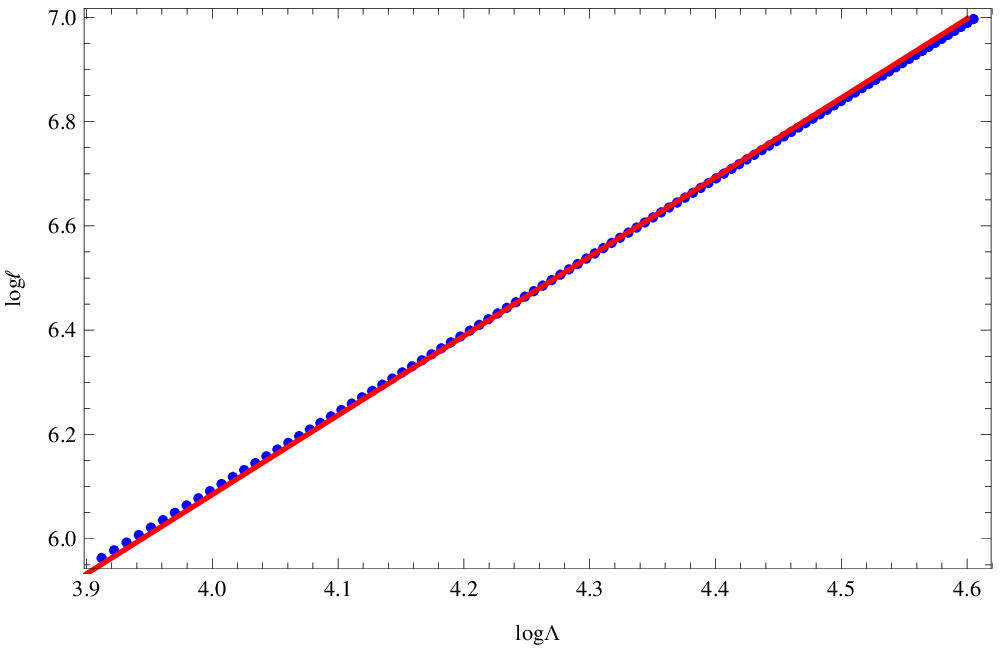}
\includegraphics[width=7.5cm]{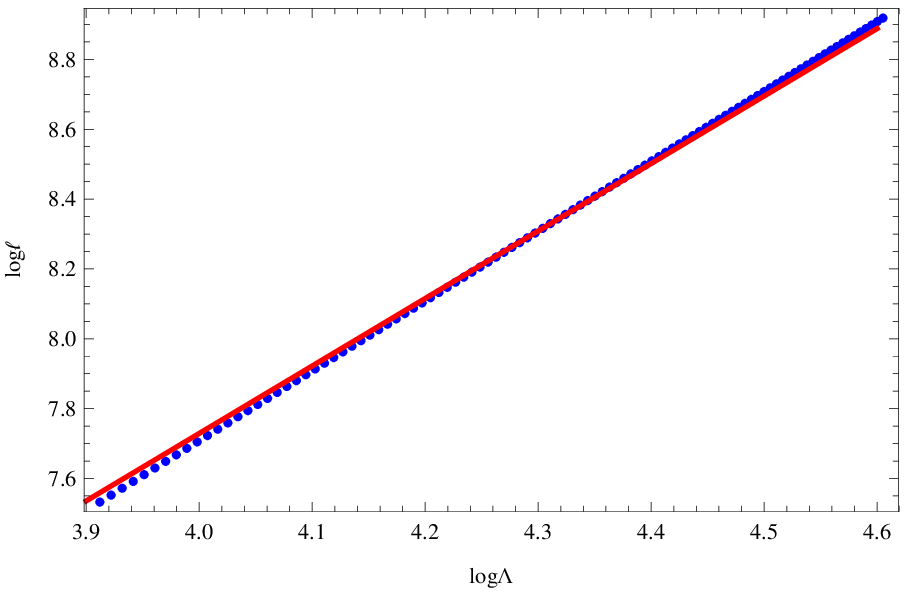}
\caption{The logarithm of $\ell$ ($\log{\ell}$) as a function of $\log{\Lambda}$. Left plot for $d=2$ and right plot for $d=3$. The blue dotted line shows the numerical result of eq.(\ref{newell}), the red line is the fitted curve.}
}
\end{figure}
\begin{figure}[tbp]{
\includegraphics[width=7.5cm]{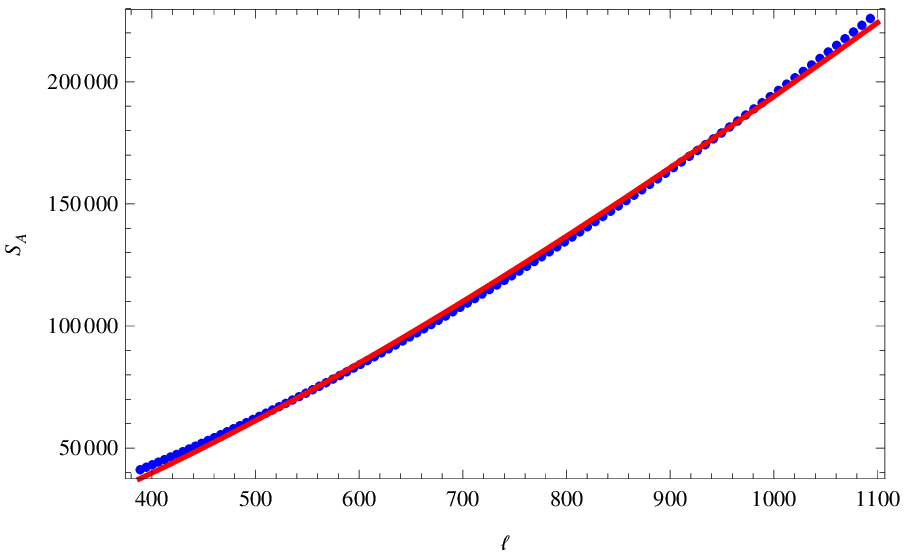}
\includegraphics[width=7.5cm]{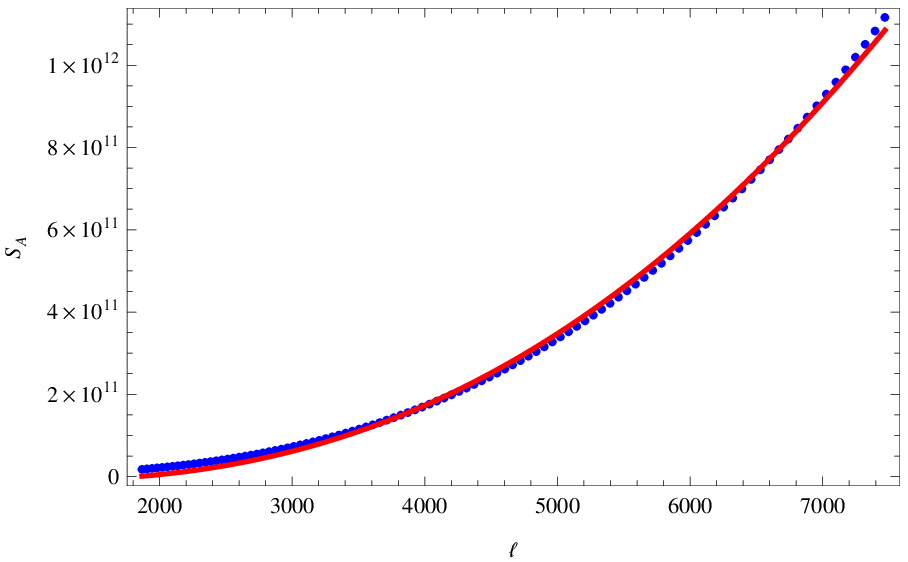}
\caption{The finite part of $S_A$ (denoted by $(S_A)_{fin}$) as a function of $\ell$. Left plot for $d=2$ and right plot for $d=3$. The blue dotted line shows the numerical result of eq.(\ref{bsasa}), the red line is the fitted curve.}
}
\end{figure}
We plot the logarithm of the width $\log{(\ell/r_F)}$ and the finite part of the entanglement entropy $(S_A)_{fin}$ in Fig.\ 1 and Fig.\ 2\footnote{As it was shown in these two plots, there appears a systematic deviation for the numerical points from the fitted curve. It behaves so because we do the numerical calculation by choosing a finite cut-off UV scale $r_{UV}\sim 10^{-2}$. As it is shown in Appendix, this will lead to some small modifications if $r_{UV}$ is not sufficiently close to the boundary.  } for $d=2\  \mbox{and}\ d=3$. In both cases, we have set $r_F=1$, $R=1$, $L=\ell$, $\Lambda \in [50,100]$ and $r_{UV}=r_F/\Lambda$. Their behaviors are well approximated by
\begin{eqnarray}
\log{\ell} &\approx& 1.52119 \log{\Lambda} \label{fit1}\\
(S_A)_{fin} &\approx& 103.09\ell \log{\ell}-518.124\ell
\end{eqnarray}
For $d=2$. And
\begin{eqnarray}
\log{\ell} &\approx& 1.93226 \log{\Lambda} \\
(S_A)_{fin} &\approx& 13838.8\ell^2 \log{\ell}-103972\ell^2  \label{fit4}
\end{eqnarray}
For $d=3$. Thus we can numerically confirm the analytical approach given above is indeed rational, as we have expected.
\section{Anisotropic systems}
In the sections above, we discussed isotropic systems with Fermi surfaces in holography by requiring the logarithmic violation of the entanglement entropy. When the size of the systems is large enough, this behavior is proposed to be general, independent of the space shapes of the systems \cite{1}. In the absence of a proof in the boundary theory, we argue that this behavior is also independent of the shape of the momentum space i.e. the shape of the Fermi surfaces. In fact, people have found this behavior by simple dimensional analysis\cite{14}. So eq.(\ref{2}) may be also effective for anisotropic systems only if we reinterpret the scale parameter $k_F$ as the average of the Fermi momentums in different directions.\\
 To put it simply, we are only interested in the two spatial dimensional systems in the following. The new metric is taken to be
\begin{equation} ds^2=\frac{R^2}{r^2} (-f(r)dt^2+g(r)dr^2+b(r)dx_1^2+a(r)dx_2^2) \label{newmetric} \end{equation}
We require that this background is also asymptotically AdS, thus
\begin{equation} f(0)=g(0)=a(0)=b(0)=1 \end{equation}
And the subsystem A is still a strip one
\begin{equation} A=\{(x_1,x_2)|-\frac{\ell}{2}\leq x_1 \leq \frac{\ell}{2},0\leq x_2\leq L\} \end{equation}
By the same procedure presented in section 1, we find
\begin{equation} x_1'=\frac{r^2}{r_*^2} \sqrt{\frac{g(r)}{a(r) b(r)^2-b(r) (\frac{r^4}{r_*^4})}} \end{equation}
 where $r_*$ is defined as the turning point of the minimal surface which makes $x_1'$ divergent, thus we have
 \begin{equation} a(r_*)b(r_*)\equiv 1  \label{ab1}\end{equation}
 To meet this condition, we first assume
 \begin{equation} a(r)b(r)\equiv 1 \label{ab}\end{equation}
 Therefore, we obtain
 \begin{equation} \ell=2 \int_0^{r_*} \mathrm{d}r \frac{r^2}{r_*^2} \sqrt{\frac{a(r)g(r)}{1-\frac{r^4}{r_*^4}}} \label{asell}\end{equation}
 \begin{equation} S_A=2R^2L \int_\epsilon^{r_*} \frac{\mathrm{d}r}{r^2} \sqrt{\frac{a(r)g(r)}{1-\frac{r^4}{r_*^4}}} \label{assa}\end{equation}
 Compare eq.(\ref{asell}) and eq.(\ref{assa}) with eq.(\ref{ell}) and eq.(\ref{sa}), we only need to replace $g(r)$ by $a(r)g(r)$  when moving from isotropic systems to anisotropic ones. Thus the argument presented in section 1 can be directly used to probe the anisotropic Fermi surfaces. From eq.(\ref{gr}) and eq.(\ref{grgr}), we find that there are several series of solutions to show the logarithmic behavior of the entanglement entropy. We list them in the following
 \begin{equation} a(r)g(r)=(\frac{r}{r_F})^2\quad;\ \mbox{IR}\ \label{42} \end{equation}
 \begin{equation} a(r)g(r)=1+\frac{r}{r_F}\quad;\ \mbox{UV}\ \label{43}\end{equation}
 \begin{equation} a(r)g(r)=(\frac{r_F}{r})^{(2d-4)} [1+(\frac{r}{r_F})^{(d-1)}]\quad;\ d\geq 3\ ;\ \mbox{UV} \label{44}\end{equation}
where IR, UV symbol denotes the cases of which part of $\gamma_A$ dominating the area integral.
Apparently, eq.(\ref{42}) and eq.(\ref{43}) is from eq.(\ref{ell}) and eq.(\ref{sa}) by setting $d=2$ while eq.(\ref{44}) needs further analysis. Substitute eq.(\ref{44}) into eq.(\ref{assa}) and compare with eq.(\ref{bsasa}), the two integral still differ by a functional factor $\sqrt{(1-u^{2d})/(1-u^4)}$, where $u=r/r_* \in [0,1]$. One can prove that this function is monotonic increasing and varys from $1$ to $\sqrt{d/2}$. For any finite $d$  which is of order $O(\Lambda^0)$, this function can be estimated by its average which contributes only a constant factor in eq.(\ref{assa}). For large $d$ limit, this estimation is however broken. But, if the $u^{2d} $ and $u^4$ terms can be omitted, we can still get eq.(\ref{44}). The point is the dominant contribution to $S_A$ is from the deep UV region. \\
Furthermore, if we always believe this point, the condition (\ref{ab}) is no longer needed. Thus, we will find
\begin{eqnarray}
S_A &=& 2R^2L\int_\epsilon^{r*}\frac{\mathrm{d}r}{r^2}\sqrt{\frac{b(r)a^2(r)g(r)}{b(r)a(r)-\frac{r^4}{r_*^4}}} \nonumber \\
 &\approx&  2R^2L\int_\epsilon^{r_*}\frac{\mathrm{d}r}{r^2}\sqrt{a(r)g(r)} \label{311sa}
\end{eqnarray}
Compare eq.(\ref{311sa}) with eq.(\ref{sasa}), we again obtain eq.(\ref{43}) and eq.(\ref{44})! The big difference is we have much more freedom to select proper functions $a(r)$ and $b(r)$ now since we only need to meet the weaken condition (\ref{ab1}) instead of eq.(\ref{ab}). For example, we can set $a(r)b(r)\equiv 1$ only holds in deep IR which is clearly weaken than eq.(\ref{ab}) but makes eq.(\ref{ab1}) satisfied automatically. \\
Note that in the case of eq.(\ref{42}) and eq.(\ref{43}) (with condition (\ref{ab})), the width and the entanglement entropy is precisely equal to the one of isotropic systems while in the case of eq.(\ref{43}) (when condition (\ref{ab})) broken) and eq.(\ref{44}) only their logarithm keep the same order of $\Lambda$. In form, eq.(\ref{44}) reduces to eq.(\ref{43}) when $d=2$ and to eq.(\ref{44}) when $d=1$ which differ by an unimportant $2$ factor. Thus eq.(\ref{44}) represents the general configuration of the minimal area surface $\gamma_A$.  \\
To specify the behavior of the specific heat, we impose null conditions on the metric (\ref{newmetric}). From equations (\ref{null})-(\ref{nu}), we get
 \begin{equation}\frac{g'}{rg}+\frac{f'}{rf}+\frac{a'^2}{2a^2}\leq 0  \end{equation}
 \begin{equation} (\frac{f''}{2f}-\frac{f'g'}{4fg}-\frac{f'}{rf}-\frac{f'^2}{4f^2})+[\frac{a''}{2a}-\frac{a'^2}{2a^2}+\frac{a'}{a} (\frac{f'}{4f}-\frac{g'}{4g}-\frac 1r)] \geq 0 \end{equation}
We only interested in the deep IR region and assume
 \begin{equation} f(r)\propto r^{-2m},\ g(r)\propto r^{2n},\ a(r)\propto r^{2p} \label{fga}\end{equation}
 which leads to
  \begin{equation} m\geq n+p^2 \label{48}\end{equation}
   \begin{equation} (m-p)(m+n+3)\geq 0 \label{49}\end{equation}
 On the other hand, according to equations (\ref{42})-(\ref{44}), the logarithmic behavior of $S_A$ requires in the deep IR
  \begin{equation} n+p=(3-d)/2\ ,\ d\geq 1  \label{irnp}\end{equation}
For physically sensible matter fields, $\alpha >0$ which is equivalent to
 \begin{equation} m+n+1>0 \label{51}\end{equation}
In the end, we obtain
 \begin{equation} d=1,\ \alpha \leq \frac{2}{(p-1)^2+2}\ ,\ -\infty <p<+\infty\ ,\ m \geq p^2-p+1\  \label{52}\end{equation}
 \begin{eqnarray}
 d=2,\ \alpha &\leq& \frac{2}{(p-1)^2+1}\ ,\  p> 1+\sqrt{2}/2\ \mbox{or}\  p<1-\sqrt{2}/2\ , m \geq p^2-p+1/2 \nonumber\\
 \alpha &\leq& 4/3\ ,\  1-\sqrt{2}/2<p<1+\sqrt{2}/2\ ,\ m\geq p\  \label{d2}
\end{eqnarray}
 \begin{eqnarray} d=3,\
 \alpha &\leq& \ \frac{2}{(p-1)^2}\ ,\  p\geq 2 \ \mbox{or}\  p\leq 0\ ,\ m\geq p^2-p\  \nonumber\\
 \alpha &\leq& 2\ ,\   0<p<2\ ,\ m\geq p\
 \label{54}\end{eqnarray}
 \begin{eqnarray} d=4\ ,
  \alpha &\leq& \frac{2}{(p-1)^2-1}\ ,\  p\geq 1+\sqrt{3/2}\ \mbox{or}\ p\leq 1-\sqrt{3/2}\ ,\ m\geq p^2-p-1/2\ \nonumber\\
  \alpha &\leq& 4\ ,\  1-\sqrt{3/2}<p<1+\sqrt{3/2}\ ,\ m\geq p\
 \end{eqnarray}
 \begin{eqnarray} d\geq 5,\
 \alpha &\leq& \frac{2}{(p-1)^2+(3-d)}\ ,\  p>p_{1+}\ \mbox{or}\ p<p_{1-}\ ,\ m\geq p^2-p+(3-d)/2\ \nonumber\\
 \alpha &<&+\infty\ ,\  p_{2-}<p<p_{2+}\ , m> p+(d-5)/2
 \label{55}\end{eqnarray}
 where $p_{1\pm}=1\pm \sqrt{d-3}$ and $p_{2\pm}=1\pm \sqrt{(d-1)/2}$. From equations (\ref{52})-(\ref{55}), one can fix $\alpha$ by selecting proper $p$ and $m$ in the permitted interval. Obviously, when $d=1$, we have $\alpha \leq 1$, thus Fermi and non-Fermi liquids can be constructed but when $d\geq 2$, even $\alpha >1$ is allowed in (\ref{d2})-(\ref{55}) which represents the thermal systems which we have no interests in. Note that every case above suggests that there exists gravity duals of Fermi and non-Fermi liquids in the purely classical limit for anisotropic systems.\\
 On the other hand, above solutions apply to $a(r)b(r)\equiv 1$ (at least in deep IR) and $a(r)$ varies with the scale parameter $r_F$ so that in the bulk interior it behaves like $a(r)\sim r^{2p}$ which is an unnecessary condition. Especially, when the anisotropy is only perturbations on the isotropic background, the behavior of $a(r)$ can be set as
 \begin{equation} a(r) = 1+\delta(a(r))=1+a_0r^{2p} \nonumber\end{equation}
 \begin{equation} \delta(a(r))/a(r) \ll 1 \ ,\ \mathrm{for\ all}\ r \label{a}\end{equation}
where $a_0$, $p$ are positive constants. Under this assumption, $a''(r)/a(r)\sim a'(r)/a(r)\sim \delta (a(r))/a(r)$ which can be neglected in the null energy conditions. Thus we obtain eq.(\ref{mn}) again. On the other hand, equations (\ref{42})-(\ref{44}) lead to
\begin{equation} n=(3-d)/2+O(\delta(a)/a) \end{equation}
Hence
\begin{eqnarray} \alpha &\leq& \frac{2}{4-d}+O(\delta(a)/a)\ ,\ d=1,2,3\ ,\ m\geq n \nonumber \\
\alpha &\leq& 4+O(\delta(a)/a)\ ,\ d=4\ ,\ m\geq 0 \nonumber \\
 \alpha &\leq& +\infty\ ,\ d\geq 5\ ,\ m>-(n+1) \label{alpha1}
\end{eqnarray}
Clearly, for $d\geq 2$ the standard Landau-Fermi liquids has been included while for $d=1$ only non-Fermi liquids is allowed.
\section{Gravity Model}
Now we would like to give an effective gravity model for systems (\ref{metric}) and (\ref{newmetric}) which have the logarithmic behavior of the entanglement entropy. The system (\ref{metric}) is isotropic and has been completely constructed by standard Einstein-Maxwell-dilaton theory \cite{1,18,19}. So we only focus on the anisotropic system (\ref{newmetric}) with $a(r)b(r)\equiv 1$.  We start from the following action
\begin{equation} S=\frac{1}{2\kappa^2} \int \mathrm{d}^{d+2}x \sqrt{-G} [\Re -2\partial_\mu{\phi}\partial^{\mu}{\phi}-V(\phi)-\frac{\kappa^2}{2}(Z_1(\phi)F^2+Z_2(\phi)H^2] \label{action} \end{equation}
where G denotes the metric determinant, $\Re$ is Ricci scalar, $V(\phi)$ and $Z_{1,2}(\phi)$ can be any function. $F$,$H$ are two Maxwell fields' strength defined by $F=dA,\ H=dB$, where $A,B$ are gauge potential. \\
The equations of motion(EOM) are
\begin{equation} G_{\mu\nu} = T_{\mu\nu} \ \label{eom1}\end{equation}
\begin{equation}\nabla_\mu{(Z_1(\phi)F^{\mu\nu})} = 0 \  \label{eom2}\end{equation}
\begin{equation}\nabla_\mu{(Z_2(\phi)H^{\mu\nu})} = 0 \  \label{eom3}\end{equation}
\begin{equation}\triangle{\phi} = \frac 14 \frac{\partial{V}}{\partial{\phi}}+\frac{\kappa^2}{8}(\frac{\partial{Z_1}}{\partial{\phi}}
F^2+\frac{\partial{Z_2}}{\partial{\phi}}H^2)\ \label{eom5} \end{equation}
where $\nabla_{\mu}$ is covariant derivative operator, $\triangle$ is Laplacian operator. $G_{\mu\nu}$ is Einstein tensor. $T_{\mu\nu}$ is energy-momentum tensor which is given by
\begin{eqnarray} T_{\mu\nu} &=& 2\partial_{\mu}{\phi}\partial_{\nu}{\phi}-g_{\mu\nu}((\partial{\phi})^2+\frac 12 V(\phi))+
\kappa^2 Z_1(\phi)(g^{\lambda\rho}F_{\mu\lambda}F_{\nu\rho}-\frac 14 g_{\mu\nu}F^2)\nonumber\\
& & +\kappa^2 Z_2(\phi)(g^{\lambda\rho}H_{\mu\lambda}H_{\nu\rho}-\frac 14 g_{\mu\nu}H^2)\  \end{eqnarray}
By definition, $G_{\mu\nu}$ of the metric (\ref{newmetric}) are found to be
\begin{eqnarray}
G_{tt}&=&-\frac{f'}{rg}(\frac{g'}{g}+\frac 3r)+\frac{f}{2g}[\frac{a''}{2a}-\frac{a'^2}{a^2}+\frac{a'}{a}(\frac{f'}{4f}-\frac{g'}{4g}-\frac 1r)]+\frac{f}{2a}\Omega(r);\nonumber \\
G_{rr} &=& -\frac{f'}{rf}-\frac{a''}{4a}-\frac{a'}{2a}(\frac{f'}{4f}-\frac{g'}{4g}-\frac 1r)+\frac{3}{r^2}-\frac{g}{2a} \Omega(r);\nonumber\\
G_{x_1x_1} &=& \frac{1}{2ga}\{[\frac{a''}{2a}+\frac{a'}{a}(\frac{f'}{4f}-\frac{g'}{4g}-\frac 1r)]+(-\frac{f'^2}{2f^2}-\frac{f'g'}{2fg}-\frac{2f'}{rf}+\frac{f''}{f}+\frac{2g'}{rg}+\frac{6}{r^2})\}-\frac{1}{2a^2}\Omega(r);\nonumber \\
G_{x_2x_2} &=& -\frac{a}{2g}\{[\frac{a''}{2a}-\frac{a''}{a^2}+\frac{a'}{a}(\frac{f'}{4f}-\frac{g'}{4g}-\frac 1r)]-(\frac{f'^2}{2f^2}+\frac{f'g'}{2fg}+\frac{2f'}{rf}-\frac{f''}{f}-\frac{2g'}{rg}-\frac{6}{r^2})\}+\frac 12\Omega(r);\nonumber \\
 & &
\label{guv}\end{eqnarray}
where $\Omega(r)$ is
\begin{equation} \Omega(r)=\frac{a'^2}{a^2}+\frac{a'}{ra}-\frac{a''}{2a}-\frac{a'f'}{4af}-\frac{a'g'}{4ag} \ ; \label{omega}\end{equation}
To realize the anisotropic metric (\ref{newmetric}), one needs $T_{\mu\nu}=T_{\mu\nu}(r)$ and $T_{x_1x_1}\neq T_{x_2x_2}$, $T_{x_1x_2}=0$\cite{15}. We search for the following solutions
\begin{equation} \phi=\phi(r)\ ,\ A=A(r)dt\ ,\ B=B(r)dx_1\  \label{abc}   \end{equation}
EOM (\ref{eom2})-(\ref{eom3}) lead to the only nonzero component of gauge fields strength is
\begin{equation} F^{rt}=\frac{Q_1}{\sqrt{-G}Z_1(\phi)}\ ,\ H^{rx_1}=\frac{Q_2}{\sqrt{-G}Z_2(\phi)}\ \label{fhl} \end{equation}
where $Q_1,Q_2$ are constants which are proportional to the corresponding conserved charge carrying by the blackholes. For convenience, we set $Q_1=Q_2=1/\kappa$. \\
Equations (\ref{eom1})-(\ref{fhl}) lead to
\begin{eqnarray}
\phi'^2 &=& \frac{g}{2a}\Omega(r)-(\frac{f'}{2fr}+\frac{g'}{2gr}+\frac{a'^2}{4a^2}+\frac 12 \Theta(r))\;\label{phi}\\
V(\phi) &=& \frac{r^2}{2g}(\frac{f'^2}{4f^2}+\frac{f'g'}{4fg}+\frac{2f'}{fr}-\frac{f''}{2f}-\frac{g'}{gr}-\frac{6}{r^2}+\Theta(r));\\
Z_{1}^{-1}(r) &=& \frac{1}{2gr^2}(\frac{f''}{f}-\frac{f'^2}{2f^2}-\frac{f'g'}{2fg}-\frac{2f'}{fr})+\frac{1}{ar^2}\Theta(r);\\
Z_{2}^{-1}(r) &=& \frac{fa}{gr^2}\Theta(r)-\frac{f}{r^2}\Omega(r);\label{z2}
\end{eqnarray}
where
\begin{equation} \Theta(r)=\frac{a''}{2a}+\frac{a'}{a}(\frac{f'}{4f}-\frac{g'}{4g}-\frac 1r)  \end{equation}
Note the dilaton equation (\ref{eom5}) remained to be a constraint. Generally, it is not satisfied automatically. Substituting (\ref{phi})-(\ref{z2}) into it will result to a constraint condition on the metric. Hence, we can't have a complete set of solution to all EOM all the time. Instead, only a subset solution is allowed. This is quiet different from the case of isotropic systems\cite{1}. To make it clear, let's consider the deep bulk interior: the metric components will behave like eq.(\ref{fga}), combined with eq.(\ref{eom5}) and (\ref{phi})-(\ref{z2}), we obtain
\begin{equation} (2m+4)p+3m+n+9 = 0\  \nonumber\end{equation}
\begin{equation} p(2p+m+3n+5)-(n+1)(m+n+3)=0\  \label{mnp} \end{equation}
 which presents a specific condition on the IR solutions of the ansatz (\ref{fga}). Only if the index $m,n,p$ locates in the restricted region, our effective gravity action (\ref{action}) does allow series of the anisotropic scaling solutions. To give the logarithmic behavior of the holographic entanglement entropy, eq.(\ref{irnp}) needs to be satisfied necessarily. Hence we can generally find three solutions for any fixed $d$ ($d\geq 1$). Unfortunately, by checking it carefully, we haven't found any solution meeting the non-trivial conditions (\ref{52})-(\ref{55}) . Thus, it seems that the gravity background (\ref{newmetric}) which has anisotropic scaling property in deep IR with the logarithmic behavior of the holographic entanglement entropy haven't been covered by above constructions.
 One possible remedial measure is when the anisotropy is only perturbations on the isotropic background, the IR solutions (\ref{fga}) will no longer valid. Instead $a(r)$ behaves like eq.(\ref{a}) in the full geometry. Since $a''(r)/a(r)\sim a'(r)/a(r)\sim \delta (a(r))/a(r)\ll 1$, we find $\Theta(r)\sim \Omega(r)\sim \delta (a(r))/a(r)\ll 1$. If ignoring this small quantity in eq.(\ref{phi})-(\ref{z2}), especially $Z_2^{-1}=0$ which leads $H$ field decoupled, the action (\ref{action}) will reduce to the standard Einstein-Maxwell-dilaton theory and so does the solutions of EOM! One can actually verify that the dilaton equation is now satisfied automatically (as emphasized in Ref.\cite{1}). On the other hand, even if keeping the first order quantity, it is also approximately satisfied, with only first order corrections at most, not broken seriously. Hence the dilaton constraint is substantially reduction. Our effective gravity model does realize systems with perturbative anisotropy, including those with Fermi surfaces absolutely.
\section{Conclusion}
In this paper, we study the hidden Fermi surfaces in holography by searching the logarithmic behavior of the entanglement entropy, without probe fermions in the bulk, which was first proposed in Ref.\cite{1}. We successfully construct the purely classical gravity duals for Fermi and non-Fermi liquids for isotropic and anisotropic systems. In both systems, the leading contribution of the finite part of the holographic entanglement entropy comes from either deep IR region or the deep UV region of the minimal area surface. It has been shown that only part of non-Fermi liquids is allowed for isotropic systems when the IR piece is dominant in Ref.\cite{1} while we explicitly prove that it is not true for generically anisotropic systems. When the deep UV region contributes to the logarithmic behavior of the entanglement entropy, both Fermi and non-Fermi liquids can be constructed in both systems. Furthermore, the hyperscaling violation exponent is found to be $\theta=d(d-1)/(d+1),\ d\geq 2$ in isotropic scaling geometries with Fermi surfaces in the UV case instead of $\theta=d-1$ in the IR case\cite{3,9,17}.
In the end, we also construct a gravity model for anisotropic background which works effectively for perturbative anisotropy and does allow series of anisotropic scaling solutions in deep bulk interior. However, the dilaton equation is not satisfied automatically and leads to a constraint on the metric. For the purpose of searching a full anisotropic solution, one needs a better gravity model. We leave it to future study.

\section{Appendix}

Let's start from the approximated expression for the entanglement entropy

\begin{equation} S_A \approx \frac{2^{d-1} R^d L^{d-1}}{r_*^{d-1}} \int_{\epsilon/r_*}^{1/\Lambda} \frac{\mathrm{d}u}{u^d} \sqrt{1+(\Lambda u)^{2n}} \label{a1}\end{equation}

where $\epsilon$ is the UV cut-off ($\epsilon\rightarrow 0$ in the UV limit), leading to the divergent part of the entanglement entropy which agrees with area law. Recall that we simply drop the $u^{2d}$ term in the denominator of eq.(\ref{bsasa}) to obtain this formula, with the assumption that the UV contribution dominant.  Since we are only interested in the leading finite part of the entanglement entropy, we will drop the UV cut-off in the lower bound of the integral. Instead we introduce a new finite UV cut-off scale parameter defined by $r_{UV}=r_F/\Lambda^{\delta}$, where $\delta$ is a positive constant to ensure $r_{UV} \ll r_F$.

Define a new variable

\begin{equation} \theta=\sqrt{1+(\Lambda u)^{2n}}  \label{a2}\end{equation}

Substitute eq.(\ref{a2}) into eq.(\ref{a1}), we deduce

\begin{equation} (S_A)_{cut}=\frac{2^{d-1} R^d L^{d-1}}{nr_F^{d-1}} \int_{\theta_1}^{\theta_2} \ \mathrm{d}\theta \frac{\theta^2}{(\theta^2-1)^{1+\frac{d-1}{2n}}} \label{a3}\end{equation}

where the integrated bounds are $\theta_1=\sqrt{1+(r_{UV}/r_F)^{2n}}$, $\theta_2=\sqrt{2}$. $(S_A)_{cut}$ denotes the entropy integral from the finite cut off scale $r_{UV}$ to $r_F$. Note that this is not the true finite part of the entanglement entropy as we will explain in the following. When $(d-1)/2n=1$, we can more explicitly write down the integrated expression as follows

\begin{equation} (S_A)_{cut}=\frac{2^{d-1} R^d L^{d-1}}{4nr_F^{d-1}} \int_{\theta_1}^{\theta_2} \ \mathrm{d}\theta [(\frac{1}{\theta-1}-\frac{1}{\theta+1})+\frac{1}{(\theta-1)^2}+\frac{1}{(\theta+1)^2}]\label{a4}\end{equation}

Now we can readily obtain the result of the integral

\begin{equation}  (S_A)_{cut}=\frac{2^{d-1} R^d L^{d-1}}{2r_F^{d-1}}\log{\frac{r_F}{r_{UV}}}+\gamma_d \frac{L^{d-1}}{r_{UV}^{d-1}}+... \end{equation}

where $\gamma_d=2^{d-1}R^d/(d-1)$, the dots denotes the small quantity of order $O(\Lambda^0)$. The first term is the intriguing logarithmic term we search but the second term is a power law which seems breaking our conclusion. In order to obtain the correct finite part of the entanglement entropy, we need to further extend the integral to the UV limit where $g(r)\approx 1$ which will lead to a corrected term $(S_A)_\epsilon$

\begin{eqnarray} (S_A)_\epsilon &\approx &\frac{2^{d-1} R^d L^{d-1}}{r_*^{d-1}} \int_{\epsilon/r_*}^{r_{UV}/r_*} \frac{\mathrm{d}u}{u^d}\nonumber\\
                            & =& \gamma_d \frac{L^{d-1}}{\epsilon^{d-1}}-\gamma_d \frac{L^{d-1}}{r_{UV}^{d-1}}
\end{eqnarray}

The total entanglement entropy is $S_A = (S_A)_{cut}+(S_A)_\epsilon $. Finally, we obtain

\begin{equation} S_A =\gamma_d \frac{L^{d-1}}{\epsilon^{d-1}}+ \frac{2^{d-1} R^d L^{d-1}}{2r_F^{d-1}}\log{\frac{r_F}{r_{UV}}}+O(\Lambda^0) \end{equation}

Evidently, by properly choosing the UV scale $r_{UV}=r_F/\Lambda^{\delta}\ll r_F$, we can exactly derive the logarithmic behavior of the entanglement entropy, as it was shown in eq.(\ref{ssa}).
\section{Acknowledgments}
I am appreciate for the JHEP referee to point out the misleading terminology ``UV-IR intermediate region" in the original manuscript. I would like to thank Professor Sije Gao, Dr. HongBao Zhang and Dr. WeiJia Li for their useful suggestions and encouragement.
I also thank Professor Norihiro lizuka for pointing out the omissive citations Ref.\cite{18,19}.
This work is supported by NSFC Grants NO.10975016 and NO.11235003.


\begin{thebibliography}{99}
\bibitem{2}J.Maldcena, Int.\ J. Theor.\ Phys.\ 38 (1999) 1113 [arXiv:hep-th/9711200]
\bibitem{3}E.Witten, arXiv:hep-th/9802150.
\bibitem{1}Noriaki Ogawa, Tadashi Takayanagi and Tomonori Ugajin, JHEP 01 (2012) 125 [arXiv:1111.1023v4 [hep-th]].
\bibitem{4}T.Faulkner, H.Liu, J.McGreevy, and D.Vegh, Phys.\ Rev.\ D  83, 125002 (2011) [arXiv:0907.2694v2 [hep-th]].
\bibitem{5}N.Iqbal, H.Liu, and M.Mezei, arXiv:1110.3814v1 [hep-th].
\bibitem{6}S.A.Hartnoll and Alireza Tavanfar, Phys.\ Rev.\ D 83, 046003 (2011) [arXiv:1008.2828 [hep-th]].
\bibitem{7}S.A.Hartnoll, D.M.Hofman, D.Vegh, JHEP 08 (2011) 096 [arXiv:1105.3197[hep-th]].
\bibitem{8}N.Iqbal an H.Liu, Class.\ Quantum Grav 29 (2012) 194004 [arXiv:1112.3671v3[hep-th]].
\bibitem{9}Liza.\ Huijse, S.Sachdev,\ B.Swinger, Phys.\ Rev.\ B 85, 035121 (2012) [arXiv:1112.0573[cond-mat.str-el]].
\bibitem{10}S.A.Hartnoll and E.Shaghoulian, JHEP 07 (2012) 078 [arXiv:1203.4236[hep-th]].
\bibitem{16}E.Shaghoulian, JHEP 05 (2012) 065 [arXiv:1112.2702v2 [hep-th]].
\bibitem{11}Shinsei Ryu and Tadashi Takayanagi, Phys.\ Rev.\ Lett 96, 181602 (2006) [arXiv:hep-th/0603001v2].
\bibitem{12}Shinsei Ryu and Tadashi Takayanagi, JHEP 08 (2006) 045 [arXiv:hep-th/0605073v3].
\bibitem{13}Tatsuma Nishioka, Shinsei Ryu and Tadashi Takayanagi, J.Phys.\ A: Math.\ Theor. 42 (2009) 504008 [arXiv:0905.0932v2 [hep-th]].
\bibitem{14}H.Liu and M.Mezei, arXiv:1202.2070v1 [hep-th].
\bibitem{18}Christos Charmousis, Blaise Gouteraux, Bom Soo Kim, Elias Kiritsis, Rene Meyer, JHEP 11 (2010) 151 [arXiv:1005.4690 [hep-th]].
\bibitem{19}Norihiro lizuka, Nilay Kundu, Prithvi Narayan and Sandip P.Trivedi, JHEP 01 (2012) 094 [arXiv:1105.1162 [hep-th]].
\bibitem{15}Norihiro lizuka and Kengo Maeda, JHEP 07 (2012) 129 [arXiv:1204.3008v1 [hep-th]].
\bibitem{17}Xi Dong, S.Harrison, S.Kachru, G.Torroba and H.Wang, JHEP 06 (2012) 041 [arXiv:1201.1905v4 [hep-th]].
\end{thebibliography}
\end{document}